\documentclass[letter]{jpsj3}

\usepackage{bm}
\usepackage{graphicx}
\usepackage{url}

\title{Maxwell's Demon in Markov Chain Monte Carlo:
\\Cooling Information Flow and Entropy Balance}

\author{Masayuki \textsc{Ohzeki}\thanks{E-mail address: mohzeki@tohoku.ac.jp}}

\inst{Graduate School of Information Sciences, Tohoku University, Sendai 980-8579, Japan\\
Department of Physics, Institute of Science, Tokyo 162-8601, Japan\\
Research and Education Institute for Semiconductors and Informatics, Kumamoto University, Kumamoto 860-8555, Japan\\
Sigma-i Co., Ltd., Tokyo 108-0075, Japan}

\abst{Markov chain Monte Carlo algorithms can be viewed as feedback devices that compare a proposed move with the target distribution and then accept or reject it. In this paper the Maxwell demon is identified with the acceptance module: it measures a proposed edge, stores the outcome in the accept/reject bit, and uses that bit to shape the probability current. The decision bit carries a genuine Shannon mutual information about the proposal, whereas only its directional part is converted into a cooling information flow. The relative-entropy relaxation rate obeys $v(t)=\dot{\mathcal I}_{\rm cool}(t)+\dot S(t)$, which separates useful cooling from housekeeping circulation in nonreversible chains.}

\kword{Markov chain Monte Carlo, Maxwell's demon, information thermodynamics, entropy production, nonreversible Markov chains}

\begin{document}
\maketitle

\textit{Introduction.}---Markov chain Monte Carlo (MCMC) methods are usually analyzed through spectral gaps, mixing times, conductance, or asymptotic variance.\cite{Levin2017,Lawler1988,Peskun1973}
These approaches are indispensable, but they describe convergence mostly as a property of a transition operator.
In an implemented algorithm, by contrast, a candidate state is generated, information about the target distribution is queried, and the move is accepted, rejected, or redirected.
This is precisely the operational structure of measurement and feedback: the algorithm acts as a Maxwell's demon that sorts proposed moves using information about the landscape.\cite{Parrondo2015,Sagawa2008,Sagawa2012}

\begin{figure*}[tb]
\begin{center}
\setlength{\unitlength}{1mm}
\begin{picture}(173,66)
\thicklines
\put(0,55){\framebox(29,10){\begin{minipage}{28mm}\centering
\scriptsize\textbf{Proposal} 
\footnotesize $q=(x,y)$
\end{minipage}}}
\put(36,55){\framebox(29,10){\begin{minipage}{28mm}\centering
\scriptsize\textbf{Measurement} 
\footnotesize $\ell(x,y)$
\end{minipage}}}
\put(72,55){\framebox(29,10){\begin{minipage}{28mm}\centering
\scriptsize\textbf{Memory}
\footnotesize $R\in\{0,1\}$
\end{minipage}}}
\put(108,55){\framebox(29,10){\begin{minipage}{28mm}\centering
\scriptsize\textbf{Feedback}
\footnotesize $A(x,y)$
\end{minipage}}}
\put(144,55){\framebox(29,10){\begin{minipage}{28mm}\centering
\scriptsize\textbf{Current}\\[-0.6mm]
\footnotesize $J_t(x,y)$
\end{minipage}}}
\put(29,60){\vector(1,0){7}}
\put(65,60){\vector(1,0){7}}
\put(101,60){\vector(1,0){7}}
\put(137,60){\vector(1,0){7}}
\put(4,32){\framebox(49,14){\begin{minipage}{47mm}\centering
\scriptsize\textbf{Acquired information} Eq.~(\ref{eq:decision_MI})\\[-0.5mm]
\footnotesize $\mathcal M_{\rm dec}=I_t(Z;R)$
\end{minipage}}}
\put(61,32){\framebox(56,14){\begin{minipage}{54mm}\centering
\scriptsize\textbf{Cooling information} Eq.~(\ref{eq:Icool})\\[-0.5mm]
\scriptsize
$\displaystyle \dot{\mathcal I}_{\rm cool}
=\frac{1}{2}\sum_{x,y}J_t(x,y)
\ln\frac{\pi(x)}{\pi(y)}$
\end{minipage}}}
\put(124,32){\framebox(49,14){\begin{minipage}{47mm}\centering
\scriptsize\textbf{Entropy balance} Eq.~(\ref{eq:balance_intro})\\[-0.5mm]
\footnotesize $v(t)=\dot{\mathcal I}_{\rm cool}(t)+\dot S(t)$
\end{minipage}}}
\put(53,39){\vector(1,0){8}}
\put(117,39){\vector(1,0){7}}
\put(86.5,55){\line(0,-1){3}}
\put(86.5,52){\line(-1,0){58}}
\put(28.5,52){\vector(0,-1){6}}
\put(158.5,55){\line(0,-1){3}}
\put(158.5,52){\line(-1,0){69.5}}
\put(89,52){\vector(0,-1){6}}
\put(127,50){\makebox(0,0){\scriptsize aligned current}}
\put(0,4){\framebox(57,17){\begin{minipage}{55mm}\centering
\scriptsize\textbf{Reversible split} Eq.~(\ref{eq:TA_split})\\[-0.1mm]
\scriptsize
$\displaystyle
\dot{\mathcal I}_{\rm cool}
=\dot{\mathcal I}_T+\dot{\mathcal I}_A$
\end{minipage}}}
\put(63,4){\framebox(49,17){\begin{minipage}{47mm}\centering
\scriptsize\textbf{Proposal fraction} Eq.~(\ref{eq:proposal_fraction})\\[-0.1mm]
\scriptsize
$F_T(0,\tau)=$
$\frac{\int_0^\tau\dot{\mathcal I}_T(t)\,{\rm d}t}
{\int_0^\tau\dot{\mathcal I}_{\rm cool}(t)\,{\rm d}t}$
\end{minipage}}}
\put(118,4){\framebox(55,17){\begin{minipage}{53mm}\centering
\scriptsize\textbf{Nonrev. split} Eq.~(\ref{eq:edge_split})\\[-0.1mm]
\scriptsize
$\dot{\mathcal I}_{\rm edge}
=\dot{\mathcal I}_{\rm cool}+\sigma_{\rm hk}$
\end{minipage}}}
\put(89,32){\line(0,-1){6}}
\put(89,26){\line(-1,0){60.5}}
\put(28.5,26){\vector(0,-1){5}}
\put(89,26){\vector(0,-1){5}}
\put(148.5,32){\vector(0,-1){11}}
\end{picture}
\end{center}
\caption{Conceptual structure of the Maxwell-demon interpretation of MCMC. The acceptance/feedback module measures the proposed edge $Z=(x,y)$ through the target log-likelihood ratio $\ell(x,y)$, records the accept/reject bit $R$, and uses this bit to create the probability current $J_t(x,y)$. Equation numbers in the boxes point to the corresponding definitions and balances in the text. The decision mutual information $\mathcal M_{\rm dec}=I(Z;R)$ quantifies acquired information, whereas $\dot{\mathcal I}_{\rm cool}$ is the part converted into cooling of the target surprisal. Reversible diagnostics split $\dot{\mathcal I}_{\rm cool}$ into proposal and acceptance contributions, while nonreversible diagnostics separate useful cooling from housekeeping circulation.}
\label{fig:demon_framework}
\end{figure*}
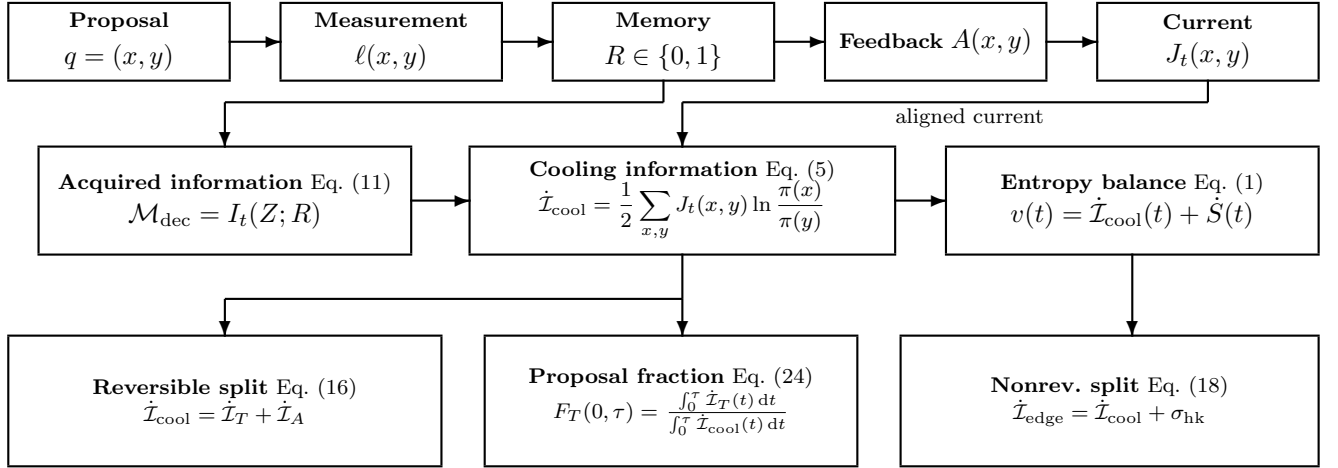

The purpose of this Letter is to make this analogy mathematically precise.
The demon is not an additional physical bath; it is the computational module that performs the cycle consisting of the measurements and feedback.
The demon is the accept/reject feedback controller: it measures the proposed edge
$q=(x,y)$ through the target log-likelihood ratio $\ell(x,y)$, stores the
decision bit $R$, and uses this bit to shape the probability current $J_t$.
There are therefore two different information quantities in the problem.
First, the accept/reject bit contains a genuine Shannon mutual information $\mathcal M_{\rm dec}=I(q;R)$ about the proposed transition.
This quantity measures how much the demon's memory knows about the candidate move.
Second, only the part of this acquired information that produces a current aligned with the target landscape is converted into relaxation.
This converted part is a current-weighted target log-likelihood, or \textit{cooling information flow} $\dot{\mathcal I}_{\rm cool}$.
For a Boltzmann target it is simply the dimensionless cooling power $-\beta \mathrm d\langle H\rangle_t/\mathrm dt$.
The exact balance derived below is
\begin{equation}
    v(t)=\dot{\mathcal I}_{\rm cool}(t)+\dot S(t),
    \label{eq:balance_intro}
\end{equation}
where $v(t)$ is the decay rate of the relative entropy to the target distribution and $S(t)$ is the Shannon entropy of the sampler.
Thus an information speed limit follows only in entropy-contracting regimes, $\dot S(t)\le 0$.
In entropy-expanding regimes, the Shannon entropy growth itself assists relaxation.
Figure~\ref{fig:demon_framework} summarizes this information-processing structure and the diagnostic quantities used below.

\textit{Relative-entropy relaxation.}---Consider a continuous-time Markov chain on a finite, irreducible state space, with transition rate $W(x|y)$ from $y$ to $x$.
For a discrete-time MCMC kernel, the same notation applies to its standard Poissonized continuous-time version; the formulas below are the infinitesimal counterpart of one-step relative-entropy contraction.
All logarithms are taken on bidirectionally connected edges with positive rates.
Let $P_t(x)$ be the distribution at time $t$, and let $\pi(x)$ be the target stationary distribution.
For equilibrium sampling, $\pi(x)=P_{\rm eq}(x)=\exp[-\beta H(x)]/Z$.
We define
\begin{equation}
    D(P_t\Vert\pi)=\sum_x P_t(x)\ln\frac{P_t(x)}{\pi(x)},\quad
    v(t)=-\frac{\mathrm d}{\mathrm dt}D(P_t\Vert\pi).
    \label{eq:KLspeed}
\end{equation}
With the probability current
\begin{equation}
    J_t(x,y)=W(x|y)P_t(y)-W(y|x)P_t(x),
\end{equation}
the master equation reads $\dot P_t(x)=\sum_y J_t(x,y)$.
The antisymmetry $J_t(x,y)=-J_t(y,x)$ gives
\begin{equation}
    v(t)=\frac{1}{2}\sum_{x,y}J_t(x,y)
    \ln\frac{P_t(y)\pi(x)}{P_t(x)\pi(y)}.
    \label{eq:nonadiabatic}
\end{equation}
This is the standard nonadiabatic entropy production rate and is non-negative for a Markov semigroup with stationary distribution $\pi$.\cite{Esposito2010}

Equation~(\ref{eq:nonadiabatic}) immediately exposes the useful information term.
Define the target surprisal potential $\Phi(x)=-\ln\pi(x)$ and the cooling information flow
\begin{align}
    \dot{\mathcal I}_{\rm cool}(t)
    &=
    \frac{1}{2}\sum_{x,y}J_t(x,y)\ln\frac{\pi(x)}{\pi(y)}
    &=
    -\frac{\mathrm d}{\mathrm dt}\sum_x P_t(x)\Phi(x).
    \label{eq:Icool}
\end{align}
The Shannon entropy of the sampler, $
    S(t)=-\sum_xP_t(x)\ln P_t(x) $
obeys
\begin{equation}
    \dot S(t)=\frac{1}{2}\sum_{x,y}J_t(x,y)
    \ln\frac{P_t(y)}{P_t(x)}.
    \label{eq:Sdot}
\end{equation}
Substituting Eqs.~(\ref{eq:Icool}) and (\ref{eq:Sdot}) into Eq.~(\ref{eq:nonadiabatic}) yields the exact balance Eq.~(\ref{eq:balance_intro}).
Equivalently, since $D(P_t\Vert\pi)=-S(t)+\langle\Phi\rangle_t$, relative-entropy relaxation is the decrease of a nonequilibrium free-surprisal.

This form corrects a common but dangerous sign intuition.
The quantity
\begin{equation}
    \dot\Sigma_{\rm Sh}(t)=
    \frac{1}{2}\sum_{x,y}J_t(x,y)\ln\frac{P_t(x)}{P_t(y)}
    =
    -\dot S(t)
    \label{eq:SigmaSh}
\end{equation}
is not an entropy production rate and has no definite sign.
Consequently,
\begin{equation}
    v(t)\le \dot{\mathcal I}_{\rm cool}(t)
    \label{eq:conditional_bound}
\end{equation}
when $\dot S(t)\le 0$.

\textit{Measurement and feedback in Metropolis-Hastings.}---We now identify where the demon's information resides in the algorithm.
For an off-diagonal Metropolis-Hastings transition, write $
    W(x|y)=T(x|y)A(x,y)
$ where $T(x|y)$ is the proposal probability and $A(x,y)$ is the acceptance probability.
The proposal step generates a candidate edge $z=(x,y)$ with probability
 $q_t(z)=P_t(y)T(x|y)$
and the demon measures the log-likelihood ratio
\begin{equation}
    \ell(x,y)=\ln\frac{\pi(x)T(y|x)}{\pi(y)T(x|y)}
    \label{eq:loglikelihood}
\end{equation}
to decide whether the candidate edge should be opened.
For the Metropolis rule $A(x,y)=\min[1,\exp \ell(x,y)]$, detailed balance is satisfied.

Let $R\in\{0,1\}$ denote the reject/accept decision for one proposal trial, with
$\mathrm{Pr}(R=1|z)=A(z)$ and mean acceptance probability
\begin{equation}
    \alpha_t=\sum_z q_t(z)A(z).
\end{equation}
For $0<\alpha_t<1$, the decision bit carries the Shannon mutual information
\begin{align}
    \mathcal M_{\rm dec}(t)
    &=
    I_t(Z;R)
    \nonumber\\
    &=
    \sum_z q_t(z)
    \left[
    A(z)\ln\frac{A(z)}{\alpha_t}
    +(1-A(z))\ln\frac{1-A(z)}{1-\alpha_t}
    \right].
    \label{eq:decision_MI}
\end{align}
This non-negative quantity is a literal information-theoretic measure: it vanishes when the accept/reject probability is independent of the proposed edge and increases when the controller selectively filters proposals using the target landscape.
It is the information stored in the demon's decision record.
Operationally, one may implement the feedback by drawing a uniform random number $U$ and setting $R=1$ when $U<A(z)$.
After averaging over $U$, the only memory of the measurement kept by the sampler is the bit $R$.
If that record is physically reset after each trial, Landauer's principle assigns an erasure cost to the memory; Eq.~(\ref{eq:decision_MI}) quantifies the part of the record correlated with the proposed edge.\cite{Sagawa2008,Sagawa2012,Horowitz2014}

The usual information-thermodynamic equalities also acquire a direct MCMC meaning.
For a trajectory $\Gamma=(x_0\to x_1\to\cdots\to x_n)$ over a time interval $[0,\tau]$, define the stochastic cooling information
\begin{equation}
    i_{\rm cool}[\Gamma]=
    \sum_{k=0}^{n-1}\ln\frac{\pi(x_{k+1})}{\pi(x_k)}
    =
    \ln\frac{\pi(x_\tau)}{\pi(x_0)}
\end{equation}
and the stochastic Shannon entropy change
\begin{equation}
    \Delta s[\Gamma]=-\ln P_\tau(x_\tau)+\ln P_0(x_0).
\end{equation}
For reversible dynamics, $i_{\rm cool}+\Delta s$ is the nonadiabatic entropy production along the trajectory and satisfies the integral fluctuation theorem
\begin{equation}
    \left\langle
    \exp[-i_{\rm cool}[\Gamma]-\Delta s[\Gamma]]
    \right\rangle=1.
    \label{eq:IFT}
\end{equation}
Jensen's inequality gives $
    \langle i_{\rm cool}\rangle+\Delta S
    =
    D(P_0\Vert\pi)-D(P_\tau\Vert\pi)
    \ge 0$
which is the finite-time version of Eq.~(\ref{eq:balance_intro}).
Thus information thermodynamics not only gives an average balance law but also constrains rare trajectories that temporarily heat the sampler or increase the relative entropy.\cite{Esposito2010,Sagawa2010,Horowitz2010}

The decision information is not identical to the relaxation speed.
Rejected moves may carry information but produce no transition current.
The part converted into relaxation is the directional current average in Eq.~(\ref{eq:Icool}): the demon cools the simulated system only when its decision record creates a net current that lowers the target surprisal $\Phi=-\ln\pi$.
Thus $\mathcal M_{\rm dec}$ measures information acquired by the demon, $\dot{\mathcal I}_{\rm cool}$ measures information actually converted into cooling, and $v$ measures the resulting decrease of relative entropy after adding the Shannon entropy change.
For a reversible chain, $W(x|y)\pi(y)=W(y|x)\pi(x)$, the edge log-ratio obeys
\begin{align}
    \dot{\mathcal I}_{\rm edge}(t)
    &=
    \frac{1}{2}\sum_{x,y}J_t(x,y)
    \ln\frac{W(x|y)}{W(y|x)}
    =
    \dot{\mathcal I}_{\rm cool}(t).
    \label{eq:Iedge_rev}
\end{align}
Thus reversible MCMC converts the antisymmetric part of the feedback likelihood ratio directly into cooling information flow.
This current average can be further split into proposal and acceptance contributions,
\begin{equation}
    \dot{\mathcal I}_{\rm cool}
    =
    \dot{\mathcal I}_{T}
    +
    \dot{\mathcal I}_{A},
    \label{eq:TA_split}
\end{equation}
with where $\dot{\mathcal I}_{T}=
    \sum_{x,y}J_t(x,y)
    \ln\left(T(x|y)/T(y|x)\right)/2$, and  $\dot{\mathcal I}_{A}=\sum_{x,y}J_t(x,y)
    \ln\left(A(x,y)/A(y,x)\right)/2$.
Equation~(\ref{eq:TA_split}) tells which part of an algorithm performs the useful sorting.
It is a decomposition of the converted information $\dot{\mathcal I}_{\rm cool}$, not of the acquired information $\mathcal M_{\rm dec}$.
For a symmetric local proposal, $\dot{\mathcal I}_{T}=0$, and cooling is carried entirely by the acceptance filter.
If many proposals are rejected, $\mathcal M_{\rm dec}$ may be large while the realized current $J_t$ is small; the demon has learned about many proposed moves but has converted little of this information into motion.
Self-learning Monte Carlo (SLMC) uses trial simulations to learn an effective model and then proposes collective moves with high acceptance probability.\cite{Liu2017}
In the present diagnostics, its advantage should appear as a larger current $J_t$, a smaller rejected part of $\mathcal M_{\rm dec}$, and, for asymmetric or learned proposal densities, a nonzero proposal contribution $\dot{\mathcal I}_{T}$.
Peskun ordering expresses the reversible limit of this idea: increasing off-diagonal transition probabilities improves asymptotic variance without changing the detailed-balance ratio, so the improvement is detected mainly through the flux magnitude rather than through a larger force.\cite{Peskun1973}

\textit{Nonreversible chains.}---When detailed balance is broken but global balance is preserved, the edge log-ratio contains both useful cooling and circulation.
Define the irreversibility force
\begin{equation}
    \mathcal F_{\rm irr}(x,y)
    =
    \ln\frac{W(x|y)\pi(y)}{W(y|x)\pi(x)}.
    \label{eq:irr_force}
\end{equation}
Then
\begin{equation}
    \dot{\mathcal I}_{\rm edge}(t)
    =
    \dot{\mathcal I}_{\rm cool}(t)+\sigma_{\rm hk}(t),
    \label{eq:edge_split}
\end{equation}
where
\begin{equation}
    \sigma_{\rm hk}(t)=
    \frac{1}{2}\sum_{x,y}J_t(x,y)\mathcal F_{\rm irr}(x,y)
    \label{eq:hk}
\end{equation}
is the housekeeping, or adiabatic, entropy production rate.\cite{Esposito2010,Hatano2001}
Accordingly, the total entropy production rate satisfies
\begin{equation}
    \sigma_{\rm tot}(t)
    =
    \dot{\mathcal I}_{\rm edge}(t)+\dot S(t)
    =
    v(t)+\sigma_{\rm hk}(t).
    \label{eq:tot_split}
\end{equation}
The housekeeping part is maintained even at stationarity, where $v(t)=0$.
It is therefore not itself relaxation speed.
The useful part is the excess, or cooling, information flow $\dot{\mathcal I}_{\rm cool}$, while $\sigma_{\rm hk}$ measures circulating information current that sustains nonequilibrium motion.
For nonreversible chains, the conversion hierarchy is enlarged to
\begin{equation}
    \dot{\mathcal I}_{\rm edge}
    =
    \dot{\mathcal I}_{\rm cool}
    +
    \sigma_{\rm hk}
    \quad\longrightarrow\quad
    v=\dot{\mathcal I}_{\rm cool}+\dot S .
    \label{eq:nr_hierarchy}
\end{equation}
Here $\dot{\mathcal I}_{\rm edge}$ is the total directed log-ratio read by the transition current, $\dot{\mathcal I}_{\rm cool}$ is the part that changes the target surprisal, and $\sigma_{\rm hk}$ is the part stored in stationary circulation.
The ratio
\begin{equation}
    \chi_{\rm cool}(t)=
    \frac{\dot{\mathcal I}_{\rm cool}(t)}
    {\dot{\mathcal I}_{\rm cool}(t)+\sigma_{\rm hk}(t)}
    =
    \frac{\dot{\mathcal I}_{\rm cool}(t)}
    {\dot{\mathcal I}_{\rm edge}(t)}
    \label{eq:chi_cool}
\end{equation}
is therefore a natural diagnostic of nonreversible algorithms whenever the denominator is positive: it measures how much of the edge information flow is used for target cooling rather than housekeeping circulation.
Good nonreversible MCMC designs make this circulation shape transient currents so that they align with the nonequilibrium force in Eq.~(\ref{eq:nonadiabatic}).
The Suwa-Todo algorithm satisfies global balance while avoiding detailed balance and is designed to reduce rejection probability.\cite{Suwa2010}
In the present language, its benefit is not an increase of the target log-ratio but an increase of useful current and a reduction of decision information wasted on rejections.
Skew detailed-balance and lifted algorithms, including irreversible Metropolis-Hastings deformations and direction- or twist-variable constructions, create persistent currents in an enlarged state space.\cite{Turitsyn2011,SakaiHukushima2013,HukushimaSakai2013,Vucelja2016}
Their advantage should be visible as a positive $\sigma_{\rm hk}$ that redirects transient currents toward the nonadiabatic force; if $\sigma_{\rm hk}$ grows without increasing $\dot{\mathcal I}_{\rm cool}$ or $v$, the irreversible circulation is mostly idle.
This interpretation is consistent with previous spectral and dynamical analyses showing that breaking detailed balance or lifting can suppress diffusive backtracking and shift slow modes of the transition operator.\cite{Chen1999,Ichiki2013,Takahashi2016_Conflict,Kaiser2017}

\textit{Numerical illustration.}---To demonstrate that the proposed quantities diagnose the mechanism of acceleration, we evaluated them exactly for a 50-state rugged ring with
$\pi(x)\propto\exp[-\beta H(x)]$.
The state space is $x=0,\ldots,49$ with periodic boundary condition, $\theta_x=2\pi x/50$, and
\begin{eqnarray}\nonumber
H(x)&=&1.6[1-\cos(2\theta_x)] \\
& & +0.55\sin(5\theta_x+0.3)+0.35\sin(9\theta_x),
\label{eq:rugged_ring}
\end{eqnarray}
shifted so that $\min_x H(x)=0$; we used $\beta=2.2$.
The initial distribution was uniform, and the master equation was integrated until $D(P_t\Vert\pi)$ was reduced by one half.
The three kernels are completely specified as follows.
The local MH kernel uses $T(x|y)=1/2$ for $x=y\pm1$ on the ring.
The learned proposal is an independence MH kernel with $T(x|y)=\tilde\pi(x)\propto\exp[-\beta_{\rm eff}H(x)]$ and $\beta_{\rm eff}=1.75$.
The irreversible kernel has nearest-neighbor rates
$W(i+1|i)=(g_i+j/2)/\pi_i$ and
$W(i|i+1)=(g_i-j/2)/\pi_{i+1}$, where
$g_i=\sqrt{\pi_i\pi_{i+1}}$,
$j=2\lambda\min_i g_i$, and $\lambda=0.82$.
Table~\ref{tab:numerics} reports the half-relaxation time $\tau_{1/2}$, the time-averaged decision information $\bar M=\tau_{1/2}^{-1}\int\mathcal M_{\rm dec}dt$, the rejection rate $\bar r$, the proposal fraction
\begin{equation}
    F_T(0,\tau)
    =
    \frac{\int_0^\tau \dot{\mathcal I}_T(t)dt}
    {\int_0^\tau \dot{\mathcal I}_{\rm cool}(t)dt}
    =
    \frac{\int_0^\tau \dot{\mathcal I}_T(t)dt}
    {\int_0^\tau [\dot{\mathcal I}_T(t)+\dot{\mathcal I}_A(t)]dt},
    \label{eq:proposal_fraction}
\end{equation}
and the finite-time cooling fraction
\begin{equation}
    \tilde\chi_{\rm cool}(0,\tau)
    =
    \frac{\int_0^\tau \dot{\mathcal I}_{\rm cool}(t)dt}
    {\int_0^\tau \dot{\mathcal I}_{\rm edge}(t)dt}
    =
    \frac{\Delta D-\Delta S}
    {\Delta D-\Delta S+\Sigma_{\rm hk}},
    \label{eq:tilde_chi}
\end{equation}
where $\Delta D=D(P_0\Vert\pi)-D(P_\tau\Vert\pi)$, $\Delta S=S(\tau)-S(0)$, and $\Sigma_{\rm hk}=\int_0^\tau\sigma_{\rm hk}(t)dt$.

\begin{table}[tb]
\caption{Exact diagnostic quantities for a 50-state rugged ring. A dash means that the corresponding accept/reject decomposition is not used. The numerical error in $v$ is less than $2\times10^{-15}$.}
\label{tab:numerics}
\footnotesize
\begin{tabular}{lccccc}
\hline
kernel & $\tau_{1/2}$ & $\bar M$ & $\bar r$ & $F_T$ & $\tilde\chi$\\
\hline
local MH & 10.13 & 0.281 & 0.300 & 0.00 & 1.00\\
learned proposal & 0.51 & 0.097 & 0.045 & 0.80 & 1.00\\
irreversible drive & 9.65 & -- & -- & -- & 0.94\\
\hline
\end{tabular}
\end{table}

The learned proposal relaxes about twenty times faster than the local Metropolis kernel.
The table shows why: the decision bit carries less information and the rejection rate is much smaller, while most of the useful cooling information is already supplied by the proposal term $\dot{\mathcal I}_T$.
This is the expected signature of SLMC-type acceleration.
The nonreversible row should be read as a diagnostic example rather than as an optimized irreversible sampler.
In this one-dimensional nearest-neighbor ring the irreversible current has little room to bypass barriers.
Nevertheless, the integrated housekeeping contribution is finite, $\Sigma_{\rm hk}=9.83\times10^{-2}$ for Table~\ref{tab:numerics}; increasing the drive to $\lambda=0.99$ gives $\tau_{1/2}=9.63$ and $\tilde\chi_{\rm cool}=0.89$, again showing circulation without a large acceleration.
Thus the near equality of the local-MH and irreversible relaxation times reflects the chosen testbed, whereas $\tilde\chi_{\rm cool}<1$ identifies the part of the edge information flow spent on housekeeping circulation rather than target cooling.
Thus the diagnostics distinguish proposal learning, acceptance filtering, and irreversible circulation in a way that conventional relaxation time alone cannot.

\textit{Conclusion.}---The Maxwell-demon interpretation of MCMC becomes sharp when one distinguishes three quantities.
The decision mutual information $\mathcal M_{\rm dec}$ is the information actually acquired by the accept/reject controller about a proposed move.
The cooling information flow $\dot{\mathcal I}_{\rm cool}$ is the part converted into a decrease of the target surprisal, equal to $-\beta\dot E$ for a Boltzmann target.
The housekeeping term $\sigma_{\rm hk}$ is the circulating information flow generated by nonreversible driving.
The exact relaxation balance is $v=\dot{\mathcal I}_{\rm cool}+\dot S$, and the Shannon term has no fixed sign.
Thus the demon does not impose a universal mutual-information ceiling on relaxation.
Rather, it uses decision information to shape probability currents, and relaxation occurs when the resulting current cools the sampler in the target landscape strongly enough to overcome any entropy contraction.
This gives a concrete information-thermodynamic design principle: maximize the conversion of decision information into cooling current while minimizing rejected or merely circulating information.
The quantities are therefore not only formal thermodynamic objects but also practical diagnostics for identifying which component of an MCMC algorithm is responsible for acceleration.

{\it Acknowledgment---.} 
The author acknowledges financial support from the Cross-ministerial Strategic Innovation Promotion Program (SIP) of the Cabinet Office (No. 23836436).

\end{document}